\documentclass[11pt,graphicx,amsmath]{article}
\usepackage{amsmath}
\usepackage{graphicx}
\usepackage{bm}
\usepackage[dvips]{color}
\usepackage{amssymb}
\usepackage{amsfonts}

\title{Initial Condition of Relic Gravitational Waves
  Constrained  by LIGO S6 and Multiple Interferometers}

\author{\small  J.W. Chen$^{1}$ \thanks{chjw@mail.ustc.edu.cn}  \ ,
                Y. Zhang$^{1}$ \thanks{yzh@ustc.edu.cn}, \, W. Zhao$^{1}$  \, M.L. Tong$^2$\\
 \small $^{1}$ Department of  Astronomy, Key Laboratory for Researches in Galaxies and Cosmology, \\
 \small    University of Science and Technology of China,   Hefei, Anhui, 230026,  China \\
 \small $^{2}$ National Time Service Center, Chinese Academy of Sciences, Xi'an,
                Shaanxi 710600,  China \\    }

 \date{}

\begin{document}
\maketitle

\def\be{\begin{equation}}
\def\ee{\end{equation}}
\def\ba{\begin{eqnarray}}
\def\ea{\end{eqnarray}}
\def\nn{\nonumber}

\baselineskip=19truept
\large

\begin{center}
\Large  Abstract
\end{center}

\begin{quote}

{The relic gravitational wave (RGW) generated during the inflation
depends on the initial condition via
 the amplitude,  the spectral index $n_t$ and the running index $\alpha_t$.
CMB observations so far have only constrained the tensor-scalar ratio $r$,
but  not $n_t$ nor $\alpha_t$.
Complementary to this,
the  ground-based interferometric detectors working at $\sim 10^2$Hz
are able to constrain the spectral indices
that influence the spectrum sensitively at high frequencies.
In this work we give a proper normalization of
the analytical spectrum at the low frequency end,
yielding  a modification by a factor of $\sim 1/50$
to the previous treatment.
We calculate the signal-noise ratios (SNR) for various ($n_t,\alpha_t$)  at fixed $r=0.2$
by S6 of LIGO H-L.
Among other things,
we  obtain the observational upper limit on the running index $\alpha_t<0.02093$
(i.e, at a detection rate $95\%$ and a false alarm rate $5\%$)
at  $(n_t=0,r=0.2)$,
 as well as a loose constraint $n_t<0.4703$ at $(\alpha _t=0,r=0.2)$.
This is consistent with the constraint
on the energy density
obtained by LIGO-Virgo Collaboration.
Extending to the four correlated detectors currently running,
the calculated  SNR  improves over  that of  LIGO H-L slightly.
When extending to the second-generation,  six correlated detectors in design,
the calculated SNR is $\sim 10^3$ times over the previous two cases,
due to the high sensitivities.
RGW can be directly detected
by the six 2nd-generation detectors
for   models with $\alpha_t>0.01364$,   or $n_t>0.2982$.

\

 }
\end{quote}


\section{Introduction}

The relic gravitational wave (RGW) is believed to
  exist in the Universe as a stochastic background
\cite{grishchuk1975,grishchuk1993,grishchuk1997,grishchuk2001,Starobinsky,Rubakov,Fabbri,Abbott,Allen1988,Allen1994,Sahnietal,Tashiro,
 Henriques2004,Henriques2010,Giovannini,Giovannini2,Maggiore,Zhao}.
The spectrum of RGW  ranges over  a very  broad frequency band $10^{-18} \sim 10^{10}$Hz,
and therefore serves as the scientific target of
various types of gravitational detections working at different frequency bands.
They includes  laser interferometers ($10^2-10^3$Hz)
such as LIGO \cite{LIGO}, Virgo \cite{VIRGO,VIRGO status}, GEO \cite{GEO,GEO intro}, KAGRA \cite{KAGRA,LCGT status,KAGRA configuration} and etc.,
and space interferometer ($\sim 10^{-3}$Hz) \cite{LISA,eLISA,LISA intro},
pulsar timing arrays ($10^{-9}-10^{-7}$Hz)    \cite{Hobbs2005,Hobbs2008,IPTA,SKA,FAST},
and CMB ($10^{-17}$Hz),
   such as  WMAP\cite{WMAP,WMAP1,WMAP7,WMAP9,WMAP9Bennett},
            Planck \cite{Planck,Planck2013}, BICEP/Keck \cite{bicep,bicep2}
                and etc.
By 2009-2010 LIGO and Virgo data,
some upper limits has been obtained on some power-law form of
the energy density $\Omega_{GW}(f)$ of
the stochastic gravitational-wave background
in relevant frequency bands \cite{Aasi2014}.

One feature of RGW is that its amplitude is higher in lower frequency bands.
The BICEP2 collaboration \cite{bicep2} has claimed the detection of
the curl-type polarization $C_l^{BB} $  of CMB
induced by RGW around  an extremely low frequency band $f\sim 10^{-17}$Hz.
However, this signal may be mixed with the thermal radiations
from galactic dust as claimed by Planck Collaboration \cite{Planck dust 2014}.
The detection, if confirmed   by other independent upcoming detections,
 will not be a surprise
since the spectrum of RGW has the highest amplitude
in this low frequency band.

After generated during the inflation,
RGW is determined by its evolution during the several stages of cosmic expansion,
as well as by the initial condition during inflation.
For the standard Big Bang model,
the  analytical solution of $h_k(\tau)$ has been obtained,
covering the whole course of evolution,
ranging from the inflation
up to the present accelerating era \cite{zhangyang2005,zhangyang2006,Miao,WangZhang,TongZhang,ZhangTong}.
Within the  cosmological model,
the slope of the spectrum,
is determined by its initial condition  during the inflation,
which, when the adiabatic vacuum is assumed,
can be summarized  into several parameters, i.e,
the spectral index $n_t$, and the running spectral index $\alpha_t\equiv dn_t/d \ln k$.
These two indices can be attributed to the detailed profile of
the potential of the  field that drives the inflation.
The over amplitude of the RGW spectrum is essentially determined by
the energy scale of the inflation.
In this work we shall improve the choice of the overall amplitude of the spectrum,
leading to a modification to our previous simple treatment   \cite{TongZhang,ZhangTong}.
The amplitude  can be also represented in terms of
the so-called tensor-scalar ratio $r$
when the scalar perturbation of the metric is known from other cosmological observations,
such as CMB and large scale structure surveys.
WMAP \cite{WMAP9} and Planck\cite{Planck}, combined with BAO, SNIa,
have put some constraints upon $r$.
In particular,
BICEP2 has given $r= 0.2$ under the default $n_t=0$.
However, since the observational data of $C_l^{BB} $ is within $l=(50\sim 150)$,
corresponding to a very narrow band,
the observational result puts little constraints on the  indices $n_t$ and  $\alpha_t$.

Nevertheless,
interferometers like LIGO and etc,
combined with the results of CMB,
can put effective constraints on $n_t$, and especially on $\alpha_t$.
This is because the profile of the RGW spectrum
is very sensitive to $n_t$ and $\alpha_t$
at high frequencies, i.e, $(10^2- 10^3)$Hz
at which LIGO and Virgo are working.
Therefore, with regard to detections of RGW,
the laser interferometers and the observation of $C^{BB}$ of CMB are complementary.
In our previous work \cite{ZhangTong},
the LIGO S5 data \cite{LIGO S5} were used to constrain $n_t$ and $\alpha_t$,
 assuming an over-estimated ratio $r=0.55$.
Now with the improved LIGO S6 data\cite{ LIGO and Virgo}
and  $r=0.2$ of BICEP2,
we shall use   the  analytical spectrum of RGW
to calculate  SNR   with the correlation between LIGO H1 and L1,
obtain an upper limits of $\alpha_t$ and $n_t$.

Correlating more detectors will provide   more accurate measurements.
By correlating
the four 1st-generation detectors, i.e,  LIGO-Hanford, LIGO-Livingston, Virgo and GEO,
using the data of sensitivity  from their scientific runs,
we calculate the optimal SNR and arrive at the detection limits of $\alpha_t$ and $n_t$,
whereby two different methods are adopted: multiple pairs and four-detector correlation.
The 2nd-generation detectors are under construction,
such as  the advanced LIGO \cite{Harry-adLIGO} and the advanced Virgo,
and KAGRA \cite{KAGRA}, AIGO \cite{AIGO},  LIGO-India \cite{LIGO-India,LIGO-India Fairhurst}.
Their sensitivities of design  will reach $\sim10^{-23}$.
We also extend the calculation to
the 2nd-generation detectors
by correlating six ones.

In section 2, we examine   the properties of the spectrum of RGW
   that are determined by the initial condition  during the inflation,
   and give a modification of its initial amplitude under the quantum normalization.

In section 3, we calculate  SNR  for various values of $\alpha_t$ and $n_t$,
                  and obtain an upper limit on $\alpha_t$,
                  based on the data of LIGO S6 and the CMB observations.

In section 4, the calculations are extended to the four detectors in operation,
        in two ways of combinations: (1) multiple pairs, and (2) four-detector correlation.

In section 5, we extend   to the  six detectors of 2nd-generation
  under construction.

In Appendix, we write down the formula of the overlapping function
         used in calculating  SNR of the pairs of detectors,
          and list the information of positions and orientations of seven detectors involved.

\section{Analytical spectrum of RGW: Normalization of Amplitude}

Consider a spatially flat RW spacetime,
\be \label{RWsynchr}
ds^2=a^2(\tau)[d\tau^2-(\delta_{ij}+h_{ij})dx^idx^j],
\ee
where $\tau$ is the conformal time,
and the perturbations $h_{ij}$ denotes RGW in the traceless and transverse gauge,
and can be written in terms of its Fourier and polarization modes
\be  \label{Fourier}
h_{ij}  ( {\bf x},\tau)= \frac{1}{(2\pi)^{3/2}}
   \int d^3k   e^{i \,\bf{k}\cdot\bf{x}}
    \sum_{s={+,\times}} {\mathop \epsilon
    \limits^s}_{ij}(k) ~ {\mathop h\limits^s}_k(\tau)
       , \,\,\,\, {\bf k}=k\hat{k},
\ee
with two polarization tensors satisfying
\[
{\mathop \epsilon  \limits^s}_{ij}(k) \delta_{ij}=0,\,\,\,\,
{\mathop \epsilon  \limits^s}_{ij}(k)  k^i=0,\,\,\,
{\mathop \epsilon  \limits^s}_{ij}(k) {\mathop \epsilon  \limits^{s'}}_{ij}(k)
       =2\delta_{ss'}.
\]
The polarizations modes ${+,\times}$ are assumed to be statistically equivalent.
In this case  the superscript $s$ can be dropped,
the evolution of
the Fourier mode $h_k(\tau)$ of RGWs is
\ba   \label{evolution}
h_k^{''}(\tau)+2\frac{a^{'}(\tau)}{a(\tau)}h_k^{'}(\tau)+k^{2}h_k(\tau)=0,
\ea
where  $k$ is the conformal wavenumber.
For each stage of  of the Big-Bang expansion, i.e,
inflation, reheating, radiation dominant, matter dominant and dark energy dominant,
the scale factor  is of a generic form $a(\tau) \propto \tau^d$
and $a^{'}/a =d/\tau$,
the solution of Eq.(\ref{evolution}) is simply a combination of two Bessel functions,
$\tau^{d-1/2} J_{d-1/2}$ and $\tau^{d-1/2}J_{-d+1/2}$.
As a prominent property of RGW, for any stage,
the solution of Eq.(\ref{evolution}) approaches to a constant,  $h_k(\tau) \sim$ const
for $k\ll a'/a$,
and takes approximately a form $h_k(\tau)\sim e^{ik\tau}/a(\tau)$ for $k\gg  a'/a$,
i.e,  the long-wavelength modes remain constant,
whereas the short-wavelength mode decrease as $1/a(\tau)$ with the cosmic expansion.
By joining these stages,
the full analytical solution $h_k(\tau)$ has been obtained,
which covers the whole course of evolution  \cite{zhangyang2005,zhangyang2006}.
Some processes occurred during the cosmic expansion,
such as neutrino free-streaming \cite{Miao},
$e^+e^-$ annihilation and QCD transition \cite{WangZhang},
 can cause small suppressions of the amplitude of RGW.
We shall not include these minor effects in this paper.
The initial condition of $h_k(\tau)$ during the inflation
will be fixed either by theoretical considerations or by actual observations.
Most of literature of RGW are
about its generation during inflation and the transition to RD stage,
and few cover the  five stages mentioned above.

The   RGW spectrum is defined by
 \be \label{defspectrum}
\int_0^{\infty}h^2(k,\tau)\frac{dk}{k}\equiv\langle0|
h^{ij}(\textbf{x},\tau)h_{ij}(\textbf{x},\tau)|0\rangle,
\ee
where $\langle 0| ... |0 \rangle$ denotes  the expectation value.
Substituting Eq.(\ref {Fourier}) into Eq.(\ref{defspectrum})
leads to the spectrum in terms of the mode $h_k(\tau)$:
\ba
h(k,\tau)=  4 \sqrt{ \pi}   k^{3/2}|h_k(\tau)|.
\label{spectrum}
\ea
When $\tau$ is set to be the present time $\tau_H$,
it gives the present spectrum,
and when $\tau$ is set to be the ending time $\tau_{1}$ of the inflation,
it will give the primordial spectrum.

To fix  the mode $h_k(\tau)$ completely,
 the initial condition has  to be specified
during the inflation,
for which the scale factor   $a(\tau) =H^{-1} |\tau|^{1+\beta}$
 with $\beta\simeq -2$ and  $H$ being the inflation rate.
We like  to emphasize that the model here
is quite general, and includes a  large class of inflation models
as long as  $a(\tau)\propto |\tau|^{1+\beta}$.
It does not necessarily rely on some scalar inflaton.
The solution for the inflationary stage   is
\ba
h_{k}(\tau) = A_k H k^{1+\beta} |x|^{-(\frac{1}{2}+\beta)}
\big[ A_1 J_{\frac{1}{2}+\beta}(x)
     +A_2 J_{-(\frac{1}{2}+\beta)}( x) \big]   ,
\label{evoveinflation}
\ea
where $x\equiv k\tau$,
two coefficients $A_1=\frac{-i}{\cos \beta\pi}\sqrt{\frac{\pi}{2}}e^{i\pi \beta/2  }$
and $A_2=i A_1 e^{-i\pi \beta}$ are
taken  so that $h_{k}(\tau) \propto H^{(2)}_{\frac{1}{2}+\beta}(k\tau)$ and
 ${\mathop {\lim} \limits_{k\rightarrow \infty}}  h_k(\tau)\propto e^{-ik \tau}$.
This choice of the initial state is also called the adiabatic vacuum \cite{Parker79}.
The coefficient factor $A_k = \frac{  \sqrt{G}}{\pi \sqrt{k} }$ is the initial amplitude
 determined by the so-called quantum normalization \cite{grishchuk97,Tong},
i.e, each $\bf k$-mode in the initial vacuum has
a zero point energy $\frac{1}{2}\hbar (\frac{2\pi k}{a})$
where  $\hbar$ is the Planck constant.
In the long wavelength limit,
the primordial spectrum takes  the following power-law form
\be \label{powerlaw}
h(k,\tau_1)= A  (\frac{k}{k_0})^{2+\beta} +O(k),
\ee
where   $A \simeq  \frac{\sqrt 2}{\pi} \frac{H}{M_{Pl}}$
  and $M_{Pl}$ is  the Planck mass,
and $k_{0}$ is a pivot conformal wavenumber  often used in CMB observations,
  corresponding to a physical wavenumber
$k_0/a(\tau_H) = 0.002 Mpc^{-1}$.
In literature \cite{WMAP1},
the leading portion of the primordial spectrum (\ref{powerlaw}) is often rewritten as
\be \label{initial}
\Delta_h (k)
=\Delta_{R} \, r^{1/2}(\frac{k}{k_{0}})^{\frac{n_t}{2}
+\frac{1}{4}\alpha_t \ln(\frac{k}{k_0})},
\ee
where
 $r\equiv \Delta^2_{h}(k_0)/\Delta^2_{R}(k_0)$ is the tensor-scalar ratio,
the tensorial spectral index $n_t=2\beta+4\equiv d \ln \Delta^2_h/d\ln k|_{k=k_0}$,
and $\alpha_t \equiv d^2 \ln \Delta^2_h/d\ln k^2 |_{k=k_0}$
is the spectral running index also introduced,
allowing  for  variations from the power-law spectrum in Eq.(\ref{powerlaw}).
Higher running indices can be further introduced \cite{KosowskyTurner},
but we are still restricted  to  only $n_t$ and  $\alpha_t$ in this paper,
as only few observational data are available currently.
Although $n_t$ and  $\alpha_t$ are formally  the coefficients
of Taylor series in terms of $\ln(k/k_0)$,
this does not necessarily mean that $\alpha_t$ would be subdominant to $n_t$
in affecting the spectrum.
In the frequency range around $f\sim 10^2$Hz for LIGO
concerned in here,
the hight of the spectral amplitude of RGW
is more sensitive to the value of $\alpha_t$ than  $n_t$, comparatively,
as Fig.\ref{hnt} and Fig\ref{halpha} demonstrate.
In our model of RGW, $r$,   $n_t $
and $\alpha_t $ by definition in (\ref{initial})  are
three independent parameters describing the primordial spectrum.
Theoretically, they are predicted by the detail of the specific inflation models.
For instance,
$r$ is proportional to $V$,
$n_t$ is a function of $V'$ (the slope of the potential),
and $\alpha_t $  is a function of  $V'$ and $V''$ (the curvature  of the potential)
in scalar  inflation models \cite{KosowskyTurner}.
For generic inflation models,
these three parameters can be independent.
Although there have been observational results and constraints
on the spectral index and running spectral index of the scalar perturbations,
so far, there is no direct observation on $r$, $n_t$  and  $\alpha_t $  of RGW  beside BICEP2.
Some very weak constraints on $r$  have  only  been given by CMB observations
   \cite{ WMAP7,WMAP9,WMAP9Bennett,Planck2013},
and this might further infer a constraint on $n_t$ via
the consistency relation within the scalar inflation models,
and different  models will give different relations such as
 $n_t=-r/8$, and its variants, etc,    \cite{LiddleLyth,Garcia-Bellido,Garriga,Bartolo,Tsujikawa}.
Most inflation modes predict the value of $n_t$ to be around $0$.
So, in absence of an observed value of $n_t$ and $\alpha_t $,
 we shall take various values of $n_t$ around $0$ in constraining  $\alpha_t $,
 and vice versa,
  in this paper.

\begin{figure}[h]
\begin{center}
\includegraphics[width=0.8\textwidth]{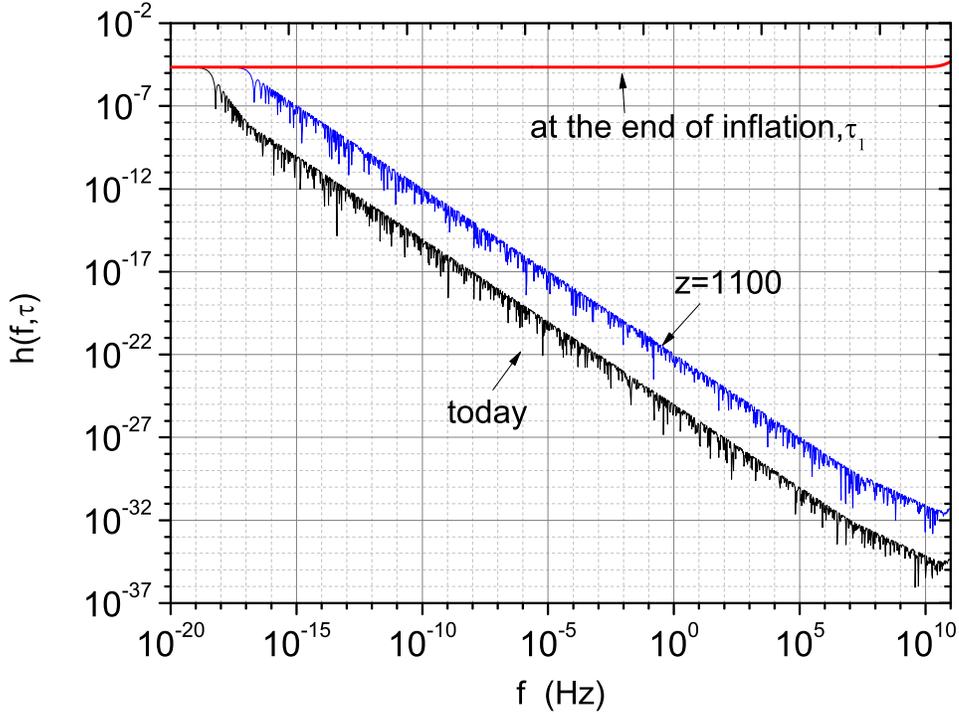}
\caption{ \label{spectrumevol}
The evolution of RGW spectrum at the fixed parameter $n_t=0$ and $ \alpha_t=0$.
The horizontal axis is the present physical frequency $f = \frac{k}{2\pi a(\tau_H)}$.}
\end{center}
\end{figure}

Fig.\ref{spectrumevol} shows the primordial spectrum $h(k,\tau_1)$ with $n_t=\alpha_t=0$
 as the top curve,
and also other two representative curves of $h(k,\tau)$ at the redshift $z=1100$
and at present, respectively.
Fig.\ref{spectrumevol}  tells the pattern of the evolution of $h(k,\tau)$,
which is decreasing with time in high frequencies.
In the expression (\ref{initial}),
$\Delta_{R}$ is the curvature perturbation at $k_0$,
which is,  by WMAP9+eCMB+BAO+$H_{0}$ observations   \cite{WMAP9Bennett},
\be \label{DeltaR}
\Delta_{R}^{2}=(2.464\pm 0.072)\times10^{-9},
\ee
and  the tensor-scalar ratio $r<0.13 \, (95\% CL)$.
Note  that the constraint $r<0.11 \, (95\% CL)$ with no scalar running,
and $r<0.26 \, (95\% CL)$ including  scalar running,
are given by Planck measurement \cite{Planck2013}.
 BICEP2   \cite{bicep2} gives
\be \label{r}
r=0.20^{+0.07}_{-0.05},
\ee
under the condition $n_t=0$ independent of the pivot scale.
The consistency relation
 would give $n_t> -0.016$ by  WMAP9 \cite{WMAP9Bennett}.
 If we were to use the consistency relation,  the BICEP2 would
 give $n_t= -0.025$.
An likelihood analysis is made on the parameters $(r,n_t)$ \cite{Suvodip},
viewing that
only with WMAP+Planck one cannot impose any constraint on $n_t$ using the value of $r$.
Ref. \cite{Suvodip} uses
the probability distribution of $r$ as the prior obtained from BICEP2 \cite{bicep2},
in combination of the likelihood contour of $(r,n_t)$ from the WMAP+Planck data,
arrive at a  probability distribution of $n_t$,
scattering around a value $n_t\sim 0.1$ loosely,
which is a poor constraint on $n_t$.
Facing this situation of absence of an observed value $n_t$,
we take ($n_t= \alpha_t=0$, $r=0.20$) as the default values of RGW,
since our model  in this paper of the initial condition of RGW is generic,
which can correspond to a large class of inflation models.
We shall compute SNR of RGW for  various values of ($n_t,\alpha_t,r$).

\begin{figure}[h]
\begin{center}
\includegraphics[width=0.8\textwidth]{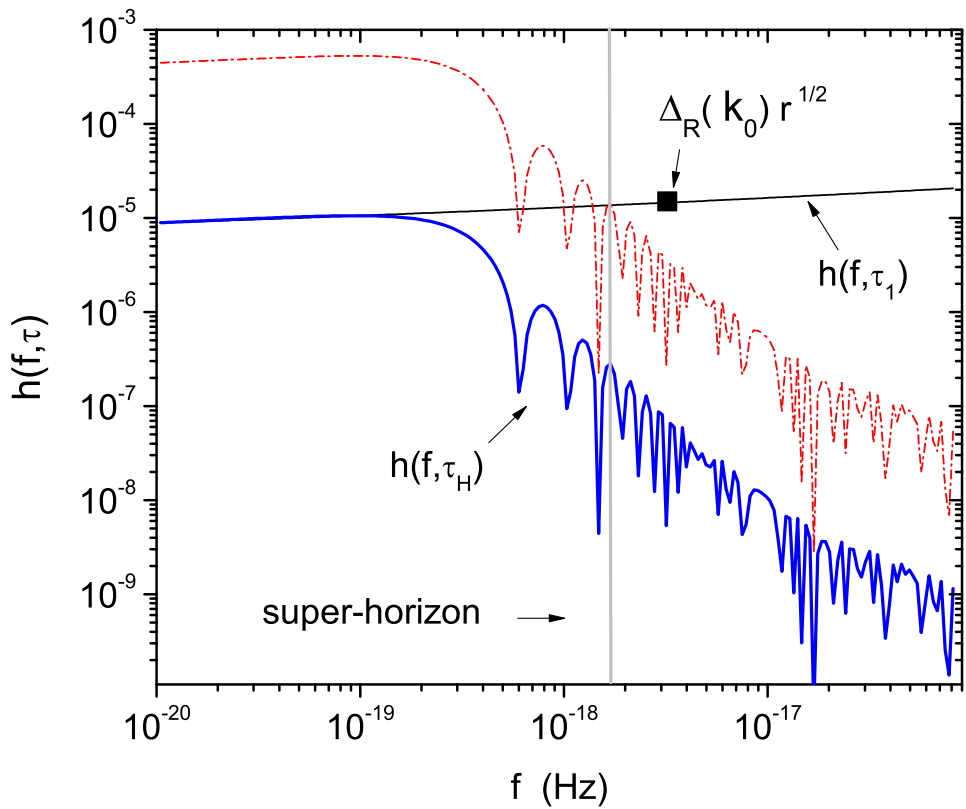}
\caption{\label{normalize}
The choice of the amplitude of the present spectrum of RGW.
Here $r=0.20, n_t=0.2, \alpha_t=0.01$ are taken for demonstration.
The above dashed curve is the overestimated spectrum in our previous work.}
\end{center}
\end{figure}

\begin{figure}[h]
\begin{center}
\includegraphics[width=0.8\textwidth]{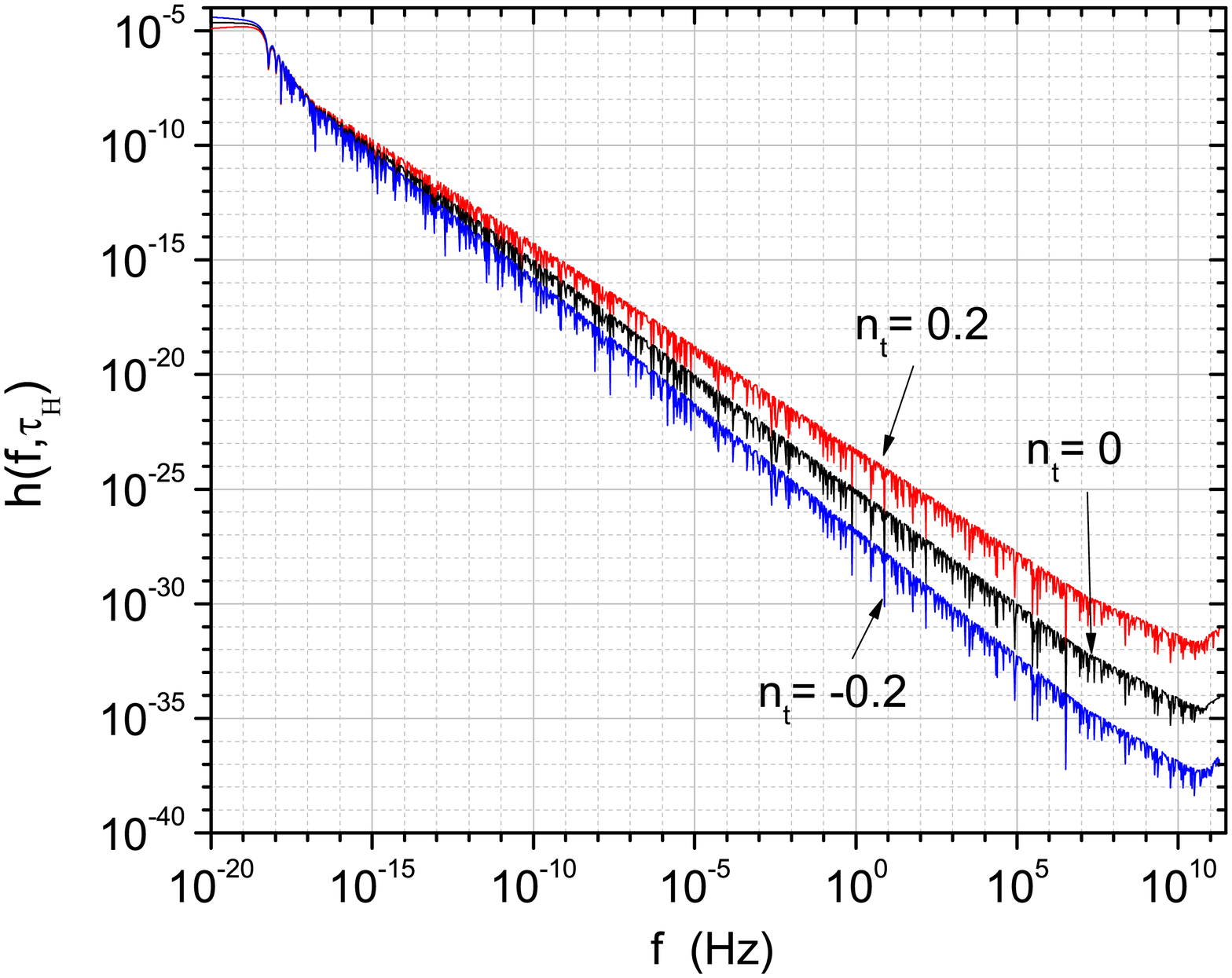}
\caption{\label{hnt}
The RGW spectrum  for various $n_t$ at  $\alpha_t=0$, $r=0.2$.}
\end{center}
\end{figure}

\begin{figure}[h]
\begin{center}
\includegraphics[width=0.8\textwidth]{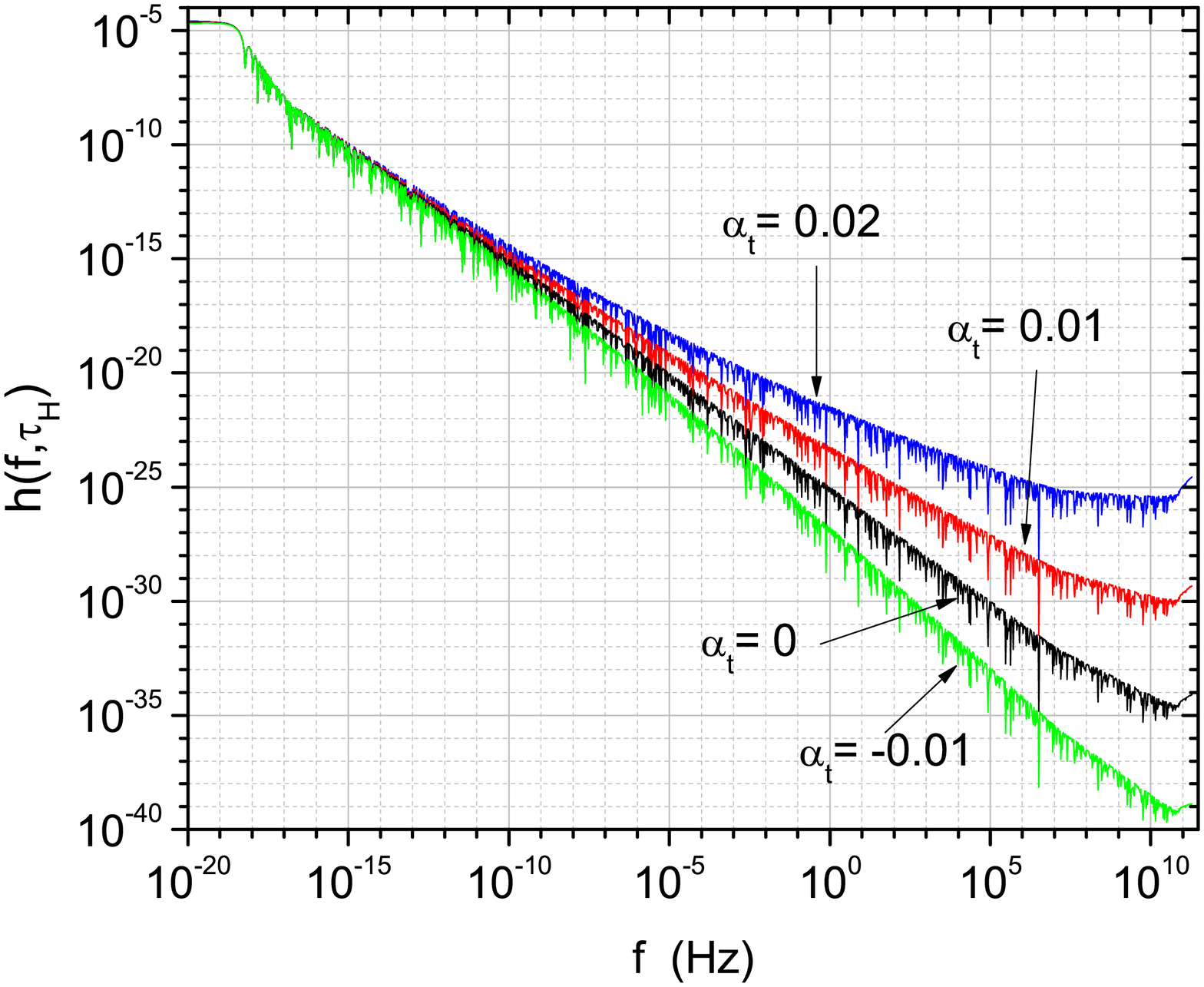}
\caption{\label{halpha}
The RGW spectrum  for  various $\alpha_t$ at $n_t=0$,  $r=0.2$.}
\end{center}
\end{figure}

As mentioned earlier,
those modes $h_k(\tau)$
remain constant  if $k\ll a'/a$ during the whole course of evolution.
Therefore, the low frequency end of the present spectrum
will remain the same as that of the  primordial spectrum, as shown in Fig.\ref{spectrumevol},
\be \label{superhorizon}
h(k,\tau_{H})=h(k,\tau_{1})  ~~~{\rm for}~~~ \frac{k}{2\pi a(\tau_H)}\ll H_0,
\ee
where
$H_0=3.24 h\times 10^{-18}$Hz is the present Hubble constant with $h\simeq 0.69$ \cite{WMAP9}.
This will tell us how to fix the undetermined amplitude of $h(k,\tau_{H})$.
We first plot the primordial $h(k,\tau_1)$  according to Eq.(\ref{initial}),
which is a roughly flat curve as shown in Fig. \ref{normalize}.
Then,  we take the low-frequency end  of
 the present spectrum  $h(k,\tau_H)$ of Eq.(\ref{spectrum})
 to be overlapping with that of the given $h(k,\tau_1)$.
This fixes the overall amplitude of $h(k,\tau_H)$.
In our previous treatment \cite{TongZhang,ZhangTong},
the amplitude at a frequency of the horizon-cross ($f \simeq H_{0}$)
was taken to be equal to the primordial value $h(k,\tau_1)$ at the same $f$.
This would lead to an overestimated amplitude (dashed line),
as Fig. \ref{normalize} shows.
Our present paper corrects it by the proper normalization,
reducing by a factor $\sim 50$.

\begin{figure}[h]
\begin{center}
\includegraphics[width=0.8\textwidth]{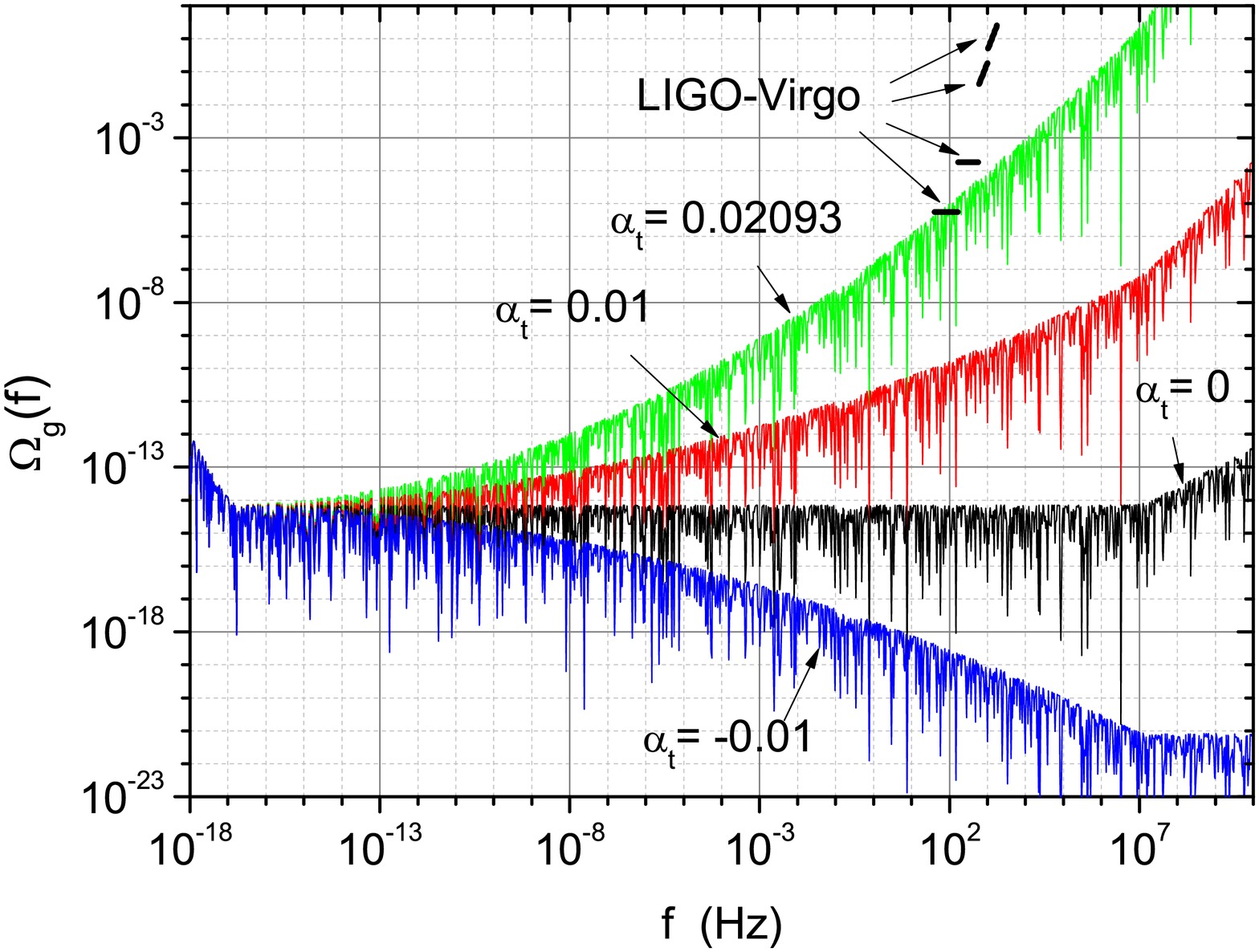}
\caption{\label{spectrumG}
The spectrum of energy density  for various $\alpha_t$
at $r=0.2$, $n_t=0$.
LIGO-Virgo data are from Ref \cite{Aasi2014}.}
\end{center}
\end{figure}

The resulting spectrum  $h(f,\tau_H)$,
drawn in Fig.\ref{hnt}  and Fig.\ref{halpha}
for various $n_t$ and $\alpha_t$,
shows
the prominent feature that it is high at low frequencies  and low at high frequencies.
At the   low frequencies $ ( 10^{-18}- 10^{-15})$Hz,
 $h(f,\tau_H) \sim (10^{-6} - 10^{-9})$
 induces  CMB anisotropy and polarization at large angles $l = (2\sim2000)$ \cite{ZhangCMB1,ZhangCMB2,ZhangCMB3,ZhangCMB4}.
In particular,
$C_l^{BB}$ detected by BICEP2 \cite{bicep2} around $l = 50\sim  150$
is induced by $h(f,\tau_H)\sim (10^{-9}\sim 10^{-8})$ in a   band
     $( 5\times 10^{-18}\sim 5\times 10^{-17})$Hz.
At the  median frequencies $ ( 10^{-9}- 10^{-7})$Hz,
 $h(f,\tau_H) \sim (10^{-17} - 10^{-16})$
 can be the target of PTA \cite{Tong}.
In this paper we are mainly concerned with  the high frequency band
            $  (10^2-10^3)$Hz for interferometers where
$h(f,\tau_H)\sim (10^{-26}\sim 10^{-27})$.
The very high frequency band around $\sim 10^{9}$Hz
can be the target of the detectors
using a polarized maser  \cite{TongZhangGaussian}.
But for $f\ge 10^{10}$Hz,
the spectrum should be subject to a regularization to subtract the divergent part of the vacuum
\cite{Parker08,Agullo08,Agullo10},
an issue not to be addressed here.

A very important property of $h(f,\tau_H)$ is that
its high frequency portion depends sensitively on the indices $n_t$ and $\alpha_t$.
A small  variation of $n_t$ and $\alpha_t$
will cause a considerable change of the amplitude at  high frequencies.
In particular, at $  10^2\sim 10^3 $Hz at which  interferometers are working,
an increase of $\alpha_t$ by   $0.01$
causes an increase of amplitude $h(f,\tau_H)$ by more than two to three orders of magnitude.
It is this property that
LIGO currently running can constrain stringently $\alpha_t$,
 and less stringently on $n_t$ comparatively,
as we shall demonstrate in the following.

The energy density of RGW is $\Omega_g= \rho_g/\rho_c$,
where $\rho_g = \frac{1}{32\pi G a^2}  h_{ij}'h'^{ij} $ and $\rho_c= 3H^2_0/8\pi G$.
The spectral energy density $\Omega_g(f)$ is defined by
$\Omega_g= \int \Omega_g(f) df/f$,
and in terms of $h(f,\tau_H)$,
 the spectral energy density is  given by
\ba \label{Omegag}
\Omega_{g}(f)  =\frac{\pi^2}{3}h^2(f,\tau_H)(\frac{f}{H_0})^2,
\label{Omega_g}
\ea
depending on the parameters $n_t$, $\alpha_t$ and  $r$ via  $h(f,\tau_H)$.
It will be used  later in Eq.(\ref{SNR}) in calculating SNR  for a pair of two detectors.
As shown in Fig.\ref{spectrumG},
for the default ($n_t= \alpha_t=0, r=0.20$),
$\Omega_{g}(f)$ is rather flat with a height $ \sim 10^{-14}$
within a large portion of frequency range.

\section{Constraint by LIGO S6}

\subsection{Constraint by a Single Detector of LIGO S6}

\begin{figure}[h]
\begin{center}
\includegraphics[width=0.8\textwidth]{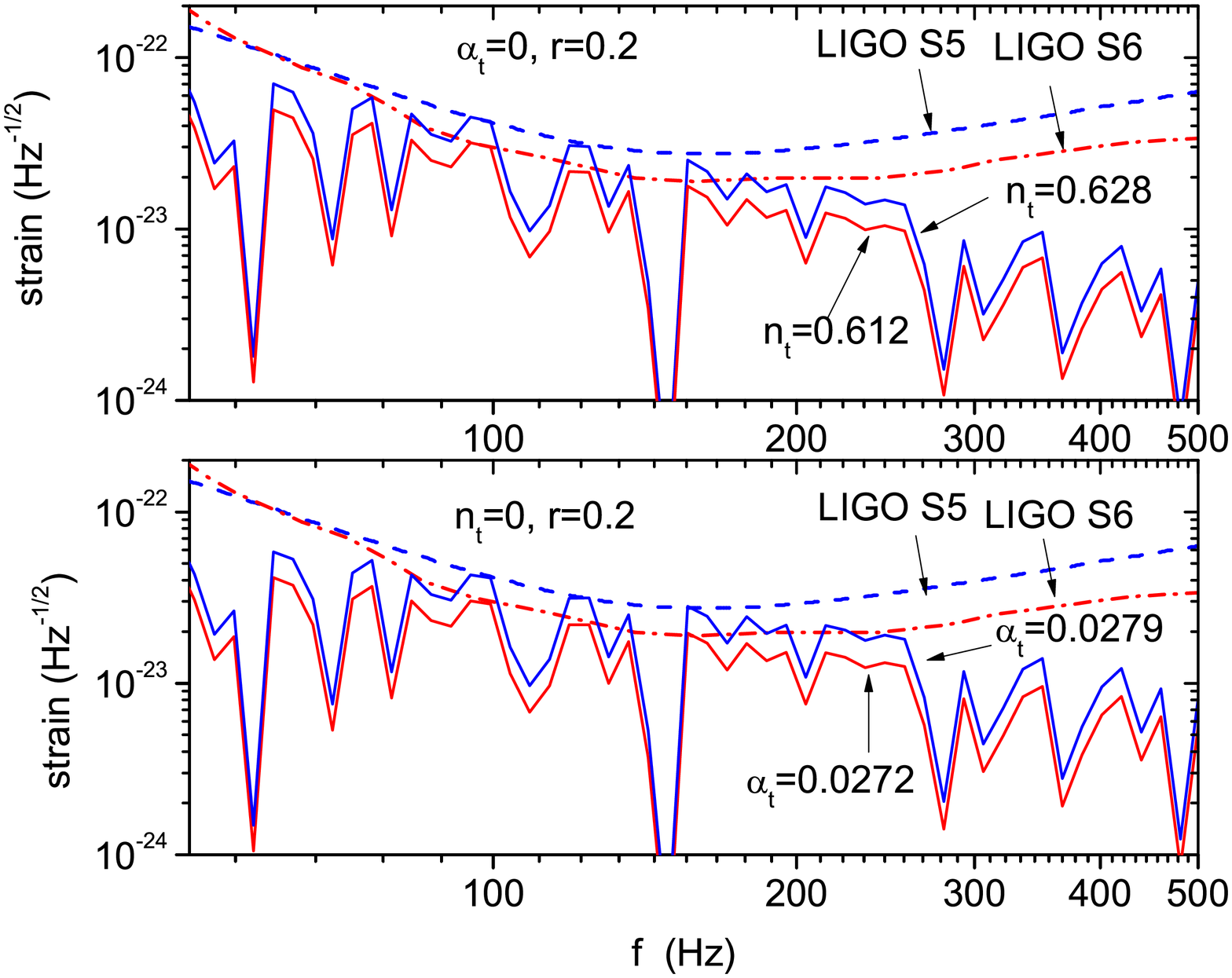}
\caption{\label{strain}
$h_c(f)\sqrt{F/2f}$ vs $\tilde{h}_f$.
This leads to constraints on RGW by   a single detector of LIGO.
Top: upper limits of $n_t$ at  $\alpha_t=0$ and $r=0.2$;
Bottom: upper limits of $\alpha_t$ at  $n_t=0$ and $r=0.2$.
}
\end{center}
\end{figure}

First let us  estimate the constraints on RGW
  by a single LIGO S6 detector running from 2009 July 7 to 2010 Oct 20 \cite{LIGO and Virgo}.
The previous work \cite{ZhangTong} did for  LIGO S5,
which was less sensitive than S6.
For  RGW  to  be detectable by a single detector
of a strain sensitivity $\tilde{h}_f$,
the simple  condition is
 \ba
h_c(f)\sqrt{F/2f}\geq \tilde{h}_f,
\label{constrain}
\ea
where $h_c(f)=h(f,\tau_H)/\sqrt{2}$ is the characteristic amplitude of RGW,
$\tilde{h}_f$ is the amplitude of noise of the detector,
which is given in Ref.\cite{LIGO and Virgo} for S6 H1 and L1 (of 2010-05),
and $F=2/5$ is the angular factor.
Although RGW has not been detected by LIGO S6,
RGW is already constrained, using Eq.(\ref{constrain}).
We plot $h_c(f)\sqrt{F/2f}$
so that it is not greater  than $\tilde{h}_f$.
This is carried out for various values of $n_t$ and $ \alpha_t$ at $ r=0.2$,
and two  specific combinations are shown in Fig. \ref{strain}.
For a comparison,  $\tilde{h}_f$ of  S5 \cite{LIGO S5} is also added into the plot.
Obviously S6 has a lower $\tilde{h}_f$  than that of S5,
yielding a more stringent constraints upon RGW.
The resulting constraint  by a single detector of LIGO S6
the resulting constraint  on the index is
\be
n_t<0.612
\ee
 at $\alpha_t=0$ and $r=0.2$,
and on the running index  is
\be \label{constralpha}
\alpha_t<0.0272
\ee
at the default $r=0.2$ and $n_t=0$.
The constraint on $n_t$ is not as stringent as that on $\alpha_t$.
This is because the amplitude $h(f,\tau_H)$
around  $(10^2-10^3)$Hz depends less sensitively to $n_t$ than to  $\alpha_t$,
{\bf as mentioned earlier.}

\subsection{Constraint   by the pair of LIGO S6}

In order to be able to detect gravitational waves,
one needs at least two correlated detectors.
Correlating two or more interferometers will increase the capability
to detect RGW
and to strengthen the constraint on $\alpha_t$.
We consider the two interferometers, LIGO H1 and L1
that have being successfully running as designed,
and calculate the SNR using the S6  data of the pair \cite{LIGO and Virgo}.
In the following, we simply denote H for H1 and L for L1.

For any two interferometer detectors $i$ and  $j$,
the formula of SNR is given by \cite{Flanagan,Allen}:
\be \label{SNR}
{\rm SNR}_{ij} =\frac{3H_0^2}{10\pi^2}\sqrt{T}
  \left[
  \int_{-\infty}^{\infty}df
  \frac{\gamma^2(f) \Omega_{g}^2(f)}
  {f^6 P_i(f) P_j  (f)}
  \right ]^{1/2},
\ee
where $H_0 $ is the Hubble constant,
$\Omega_{g}(f)$ is given in Eq.(\ref{Omegag}) from our calculation,
depending on ($n_t$, $\alpha_t$,  $r$),
and $P_i(f)= |\tilde{h}_f |^2$ is the power spectral density of the noise of the detector $i$.
For S6, we take $\tilde{h}_f$ from Ref.\cite{LIGO and Virgo}.
$T$  is the duration of detection,    $1.282$ years for LIGO S6.
$\gamma(f)$ is the overlap reduction function between two detectors,
depending on their   positions and directions.
Our calculation of $\gamma(f)$ is attached
in Appendix for the relevant pairs of
seven detectors considered in this paper.

Table \ref{S5 and S6}  shows our integration result of  SNR for S6 of LIGO H-L.
For comparison,  we also calculate for  S5  \cite{LIGO S5}.
It is seen  that the detection capability of S6 improves over that of  S5.
We mention that
in our previous work \cite{ZhangTong}, an overestimated amplitude of RGW was used.
\begin{table}[h]
\caption{\label{S5 and S6}
SNR$_{\rm HL}$  for various $\alpha_t, n_t$ at $r=0.2$.
LIGO S5 is also listed for a comparison.
}
\begin{center}
\begin{tabular}{|c|c|c|c|c|c|}
\hline
$\alpha_t,~n_t$ & -0.005,~-0.1 & 0,~0 & 0.005,~0.1 & 0.01,~0.2 & 0.015,~0.2\\
\hline
\hline
S5 & $1.6\times 10^{-13} $ & $1.9\times 10^{-9}$&$2.3\times 10^{-5} $ & $3.0\times 10^{-1} $ & $4.4\times 10^{1}$\\
S6 & $1.6\times 10^{-13} $ & $2.2\times 10^{-9}$&$3.1\times 10^{-5} $ & $4.2\times 10^{-1} $ & $6.6\times 10^{1}$\\
\hline
\end{tabular}
\end{center}
\end{table}

A constraint on $\alpha_t$ can be also given from these data of the LIGO pair.
Since SNR is implicitly a function of $\alpha_t$ via
the integrand $\Omega_{g}(f)$ at fixed $n_t$ and $r$.
Given a value of SNR,
a corresponding  $\alpha_t$ will follow from Eq.(\ref{SNR}).
In fact,
in the Frequentist approach of statistic,
SNR depends on the false alarm rate $\alpha$ and the detection rate $\gamma$
that we require for a detection.
Explicitly, the relation  is  (Eq.(4.36) of  \cite{Allen}):
\be
\label{alarm rate}
{\rm SNR}_{\alpha,\gamma}^2 = 2
\left(   {\rm  erfc}^{-1}(2\gamma)-
        {\rm erfc}^{-1}(2\alpha)   \right )^2,
\ee
where $\rm erfc^{-1}$ is the inverse complementary error function.
This relation of Eq.(\ref{alarm rate})  holds regardless of RGW.
On the other hand,
if the calculated  SNR of RGW by Eq.(\ref{SNR}) is greater than SNR$_{\alpha,\gamma}$,
the RGW signal can be effectively detected at the rates $\alpha$ and $\gamma$ as we require.
Let us take $\alpha=0.05$ and $\gamma=0.95$ for example.
Then Eq.(\ref{alarm rate}) yields a value SNR$_{\alpha,\gamma}=3.30$.
Since RGW has not yet detected by LIGO S6,
the  calculated SNR of RGW is then less than  $3.30$.

\begin{figure}[h]
\begin{center}
\includegraphics[width=0.8\textwidth]{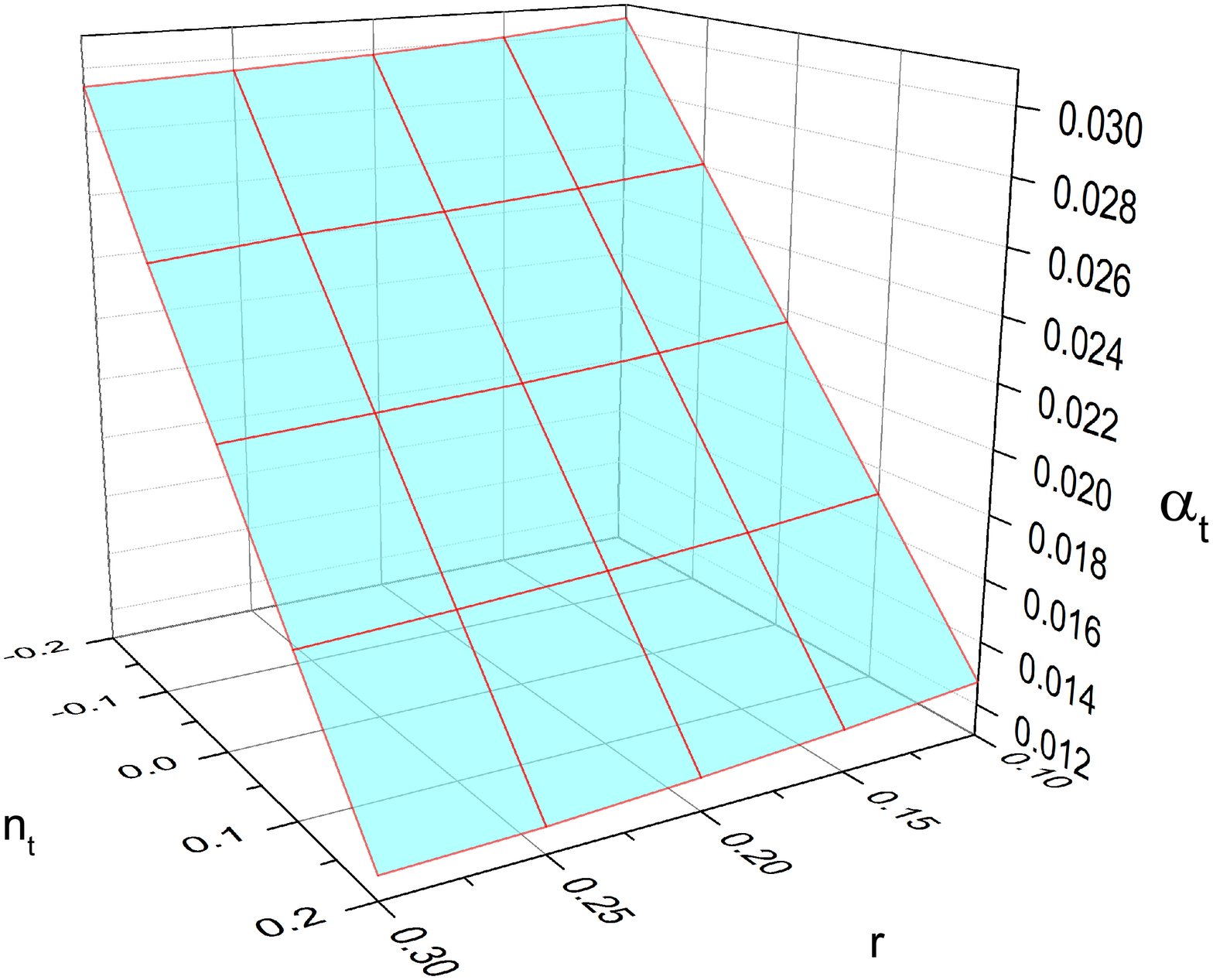}
\caption{\label{2r1}
The upper limits of $\alpha_t$ on the parameter plane $(n_t, r)$,
allowed by the observational data of the pair of LIGO S6,
to illustrate Table \ref{S6 constraint}.}
\end{center}
\end{figure}

Thereby,
a constraint on  $n_t$ can be given at fixed $\alpha _t$ and $r$.
The result is shown  in Table \ref{S6 constraint nt}.
Inflation models not satisfying the constraints will be ruled out.
In particular,
at   $\alpha _t=0$ and r=0.2,    S6 of LIGO pair on the   index is
\be
n_t< 0.4703,
\ee
which is a loose but direct constraint,
consistent with those from CMB observations, listed below Eq.(\ref{r}).
Similarly, a constraint on  $\alpha_t$ can be given at fixed $n_t$ and $r$.
shown  in Table \ref{S6 constraint}  and Fig.\ref{2r1}.
And at   $n_t=0$ and r=0.2,
the constraint of S6 of LIGO pair on the running index is
\be \label{constr2}
\alpha_t<0.02093,
\ee
more stringent than that in Eq.(\ref{constralpha}) of a single detector.
Table \ref{S6 constraint}  is the first constraint on the running index  $\alpha_t$
based on actual  observations.
The observations of CMB  so far have no constraint on $\alpha_t$
because the large angular data of CMB are not very sensitive to $\alpha_t$.
This also shows the advantage of interferometers
as a complementary to those CMB observations.

We notice that
LIGO and Virgo \cite{Aasi2014} use the data of 2009-2010 and
give the constraints
on the total energy density of
a general stochastic gravitational-wave background,
at four different frequency bands.
These  are also plotted in Fig. \ref{spectrumG} to compare with our constraints.
As Fig. \ref{spectrumG} shows clearly,
in particular,
their constraint $\Omega_{GW}(f) < 5.6\times 10^{-6}$ in $(41-169)$Hz is
consistent with our constraint in Eq.(\ref{constr2}).
However, the other three upper limits of theirs are quite higher than
our $\Omega_g(f)$ for the case  $\alpha_t= 0.02093$.
It is remarked that the constraints on the specific parameters of RGW
have been made possible
with the help of the analytical solution of RGW.

\begin{table}[h]
\caption{\label{S6 constraint nt}
The upper limits of $n_t$ constrained by the pair of LIGO   S6,
at the false alarm rate $\alpha=5\%$ and the detection rate $ \gamma=95\%$.}
\begin{center}
\begin{tabular}{|l|c|c|c|c|c|}
\hline
   & $\alpha_t=-0.01$ & $\alpha_t=0$ & $\alpha_t=0.01$ & $\alpha_t=0.02$ & $\alpha_t=0.03$ \\
\hline
\hline
$r=0.1$ & 0.7103 & 0.4858 & 0.2613 & 0.0363 & -0.1886 \\
$r=0.15$ &  0.7012 & 0.4769 & 0.2522 & 0.0273 & -0.1976 \\
$r=0.2$ &  0.6949 & 0.4703 & 0.2457 & 0.0209 & -0.2039 \\
$r=0.25$ &  0.6899 & 0.4653 &0.2407 & 0.0160 & -0.2089 \\
$r=0.3$ & 0.6858 & 0.4612 & 0.2367 & 0.0120 & -0.2129\\
\hline
\end{tabular}
\end{center}
\end{table}
\begin{table}[h]
\caption{\label{S6 constraint}
The upper limits of $\alpha_t$ constrained by the pair of LIGO   S6,
at the false alarm rate $\alpha=5\%$ and the detection rate $ \gamma=95\%$.}
\begin{center}
\begin{tabular}{|l|c|c|c|c|c|}
\hline
   & $n_t=-0.2$ & $n_t=-0.1$ & $n_t=0$ & $n_t=0.1$ & $n_t=0.2$\\
\hline
\hline
$r=0.1$ & 0.03051 & 0.02606 & 0.02162 & 0.01717 & 0.01272\\
$r=0.15$ & 0.03011 & 0.02566 & 0.02122 & 0.01677 & 0.01232\\
$r=0.2$ & 0.02982  & 0.02538 & 0.02093 & 0.01648 & 0.01203\\
$r=0.25$ & 0.02960 & 0.02516 & 0.02071 & 0.01626 & 0.01181\\
$r=0.3$ & 0.02942 & 0.02478 & 0.02053 & 0.01608 & 0.01173 \\
\hline
\end{tabular}
\end{center}
\end{table}

\section{Detection  by four  detectors of 1st-generation}
More detectors correlated have a higher capability
to directly detect RGWs.
Now we calculate the SNRs of four correlated detectors:
LIGO-Hanford(H), LIGO-Livingston(L) and Virgo(V), Geo(G).
These detectors  have carried out the  scientific runs,
and their  data of sensitivities $ |\tilde{h}_f |$  are available.
For H and L, we use LIGO S6 data;
and for V, we use the VSR2 data.
These three sets data are reported in Fig.1 of reference \cite{LIGO and Virgo}.
For Geo,  we use the Geo600 sensitivity \cite{GEO}.
Although these four detectors actually  run at different time and over different durations,
for simplicity of our calculation,
we suppose they were  running at the same time
and for the same duration  $T=1.282 $ years as  LIGO S6.
There are two ways to combine 4 detectors to sum up the total of
the optimal SNR \cite{Allen}:
(1) multiple pairs;
(2) four-detector correlation.

\subsection{multiple pairs}
Four detectors can be grouped into 6 pairs, each pair consists of 2 correlated detectors,
and there is no correlation between different pairs. For the four detectors of G, H, L and V,
the squared optimal SNR of    multiple pairs is (Eq.(5.46) in Ref.\cite{Allen})
\be \label{multiple pairs}
{\rm   SNR^2_{optimal} = SNR^2_{GH}+  SNR^2_{GL} + SNR^2_{GV}
                 + SNR^2_{HL}+ SNR^2_{HV}+ SNR^2_{LV}  },
\ee
where SNR$_{ij}^2$ of each pair on the right hand side
will be  calculated by the formula in Eq.(\ref{SNR}).
Eq.(\ref{multiple pairs}) indicates that
  SNR$ ^2_{\rm optimal}$  of multiple pairs is
always greater than the SNR$_{ij}$ of any   pair.
Using the overlapping function $\gamma(f)$ and
the relevant information of the positions and orientations
in Appendix,
we calculate each SNR$_{ij}^2$ involved in Eq.(\ref{multiple pairs}),
yielding  SNR$ ^2_{\rm optimal}$.
The result is shown in Table \ref{multiple pair SNR}.
It is seen that
SNR$ ^2_{\rm optimal}$ is greater than SNR$_{\rm HL}$ for the LIGO H-L pair
in Table  \ref{S5 and S6}  by only about $10\% \sim 20 \%$.
This is because the main contribution to SNR$ ^2_{\rm optimal}$
is from LIGO H-L pair, which has the highest sensitivities.

\begin{table}[h]
\caption{\label{multiple pair SNR}
SNR$_{\rm optimal}$  of the multiple pairs from four detectors at $ r=0.2$.
}
\begin{center}
\begin{tabular}{|c|c|c|c|c|c|}
\hline
$\alpha_t,~n_t$ & -0.005,~-0.1 & 0,~0 & 0.005,~0.1 & 0.01,~0.2 & 0.015,~0.2\\
\hline
\hline
SNR & $1.9\times 10^{-13} $ & $2.6\times 10^{-9}$&$3.5
\times 10^{-5} $ & $4.8\times 10^{-1} $ & $7.5\times 10^{1}$\\
\hline
\end{tabular}
\end{center}
\end{table}

Similar to the last section,
by requiring SNR$ _{\rm optimal} >$ SNR$_{\alpha, \gamma}=3.30$,
the multiple pair can directly detect RGW for certain value of $\alpha_t$,
as shown in Table \ref{4 multiple pair r}.
The minimum $\alpha_t $ required to be detected by 4 detectors
is lower than the constraint of LIGO pair S6, comparing to Table \ref{S6 constraint}.
In particular, for $n_t =0 $ and $r=0.2$,
 the minimum  $\alpha_t=0.02080$ for the multiple pairs is
smaller than the constraint $\alpha_t< 0.02093$ for  S6 of LIGO pair.

\begin{table}[h]
\caption{\label{4 multiple pair r}
The minimum $\alpha_t$ required to detect by  the multiple pairs from four detectors.}
\begin{center}
\begin{tabular}{|l|c|c|c|c|c|}
\hline
     & $n_t=-0.1$ & $n_t=0$ & $n_t=0.1$ \\
\hline
\hline
$r=0.1$  & 0.02593 & 0.02149 & 0.01704 \\
$r=0.15$  & 0.02553 & 0.02109 & 0.01664 \\
$r=0.2$   & 0.02525 & 0.02080 & 0.01635\\
$r=0.25$  & 0.02503 & 0.02058 & 0.01613 \\
$r=0.3$  & 0.02485 & 0.02040 & 0.01595 \\
\hline
\end{tabular}
\end{center}
\end{table}

\subsection{four-detector correlation}
Four detectors can also be treated as  a group of 4 correlated detectors,
the corresponding squared optimal SNR is   (Eq.(5.91) in \cite{Allen})
\be\label{4-detector}
{\rm SNR^2_{optimal} = SNR^2_{GH} SNR^2_{LV}+ SNR^2_{GL} SNR^2_{HV}+ SNR^2_{GV} SNR^2_{HL} }.
\ee
In this case,
the optimal SNR$_{\rm optimal}$ is quadratic in SNR$_{ij}$,
so that $SNR_{\rm optimal}$  can be enhanced greatly
if the individual SNR$_{ij}\gg 1$.
The resulting optimal SNR$_{\rm optimal}$   for the four-detector correlation
are shown in Table \ref{4-detector SNR}.

\begin{table}[h]
\caption{\label{4-detector SNR}
SNR$_{\rm optimal}$  of  the four-detector correlation at $r=0.2$.
}
\begin{center}
\begin{tabular}{|c|c|c|c|c|c|}
\hline
$\alpha_t,~n_t$ & -0.005,~-0.1 & 0,~0 & 0.005,~0.1 & 0.01,~0.2 & 0.015,~0.2\\
\hline
\hline
SNR & $4.8\times 10^{-28} $ & $1.0\times 10^{-19}$&$2.1
\times 10^{-11} $ & $4.8\times 10^{-3} $ & $1.3\times 10^{2}$\\
\hline
\end{tabular}
\end{center}
\end{table}

Similarly,
requiring SNR$_{\rm optimal} >$ SNR$_{\alpha, \gamma}=3.30$,
yields the minimum  $\alpha_t$, shown in Table \ref{4-detector r}.

\begin{table}[h]
\caption{\label{4-detector r}
The minimum  $\alpha_t$ required to be detected by the four-detector correlation}
\begin{center}
\begin{tabular}{|l|c|c|c|c|c|}
\hline
 & $n_t=-0.1$ & $n_t=0$ & $n_t=0.1$ \\
\hline
\hline
$r=0.1$ & 0.02714 & 0.02272 & 0.01830 \\
$r=0.15$  & 0.02675 & 0.02233 & 0.01791\\
$r=0.2$  & 0.02647 & 0.02205 & 0.01763 \\
$r=0.25$  & 0.02625 & 0.02183 & 0.01741 \\
$r=0.3$  & 0.02607 & 0.02165 & 0.017123  \\
\hline
\end{tabular}
\end{center}
\end{table}

\newpage
\section{Detection  by six 2nd-generation detectors}
The second-generation interferometers are currently under construction,
their goal sensitivities will eventually  reach  $10^{-24}$,
improving the sensitivity over the current detectors by more than an order.
Correlating six 2nd-generation detectors will drastically increase
SNRs and yield more stringent constraints on $\alpha_t$.
We shall  correlate six detectors:
 LIGO-Hanford(H),  LIGO-Livingston(L),  Virgo(V)  \cite{adv LIGO and Virgo},
 Kagra(K) \cite{KAGRA},  AIGO(A)  \cite{AIGO},  LIGO-India(I).
 LIGO-India  will adopt the designed spectra of sensitivity of
the advanced LIGO \cite{LIGO-India}.
In our calculation,
the BRSE mode for Kagra will be used since it has two operation modes.
Also, $T=1.282 $ years  is assumed.
From these six detectors one can construct fifteen correlated pairs.
For each pair, Eq.(\ref{SNR}) is used to calculate SNR.
The calculated result is given in Table \ref{15SNR1}.
Typically, for a pair of this kind of detectors,
SNR $\sim 10^{-6}$ for RGW of the default parameters,
and the advanced LIGO H-L will enhance the capability by thousands of times
than S6 in Table \ref{S5 and S6}.

For six detectors there are also two ways of combination:
  multiple pairs
and   six-detector correlation  \cite{Allen}.
\begin{table}[h]
\caption{\label{15SNR1}
SNR of fifteen pairs of 2nd generation detectors at $ r=0.2$.
}
\begin{center}
\begin{tabular}{|c|c|c|c|c|c|}
\hline
$\alpha_t,~n_t$ & -0.005,~-0.1 & 0,~0 & 0.005,~0.1 & 0.01,~0.2 & 0.015,~0.2\\
\hline
\hline
A-H &$2.3\times 10^{-10} $ & $2.0\times 10^{-6}$ & $ 1.9
\times 10^{-2} $& $1.7\times 10^{2} $& $2.1\times 10^{4}$\\
A-I &$1.9\times 10^{-10} $ & $1.6\times 10^{-6}$ & $ 1.4
\times 10^{-2} $& $1.2\times 10^{2} $& $1.5\times 10^{4}$\\
A-K &$3.8\times 10^{-11} $ & $3.5\times 10^{-7}$ & $ 3.5
\times 10^{-3}$& $3.5\times 10^{1} $& $4.5\times 10^{3}$\\
A-L &$2.4\times 10^{-10} $ & $2.1\times 10^{-6}$ & $ 1.9
\times 10^{-2}$& $1.8\times 10^{2} $& $2.2\times 10^{4}$\\
A-V &$4.4\times 10^{-11} $ & $4.3\times 10^{-7}$ & $ 4.4
\times 10^{-3}$& $4.5\times 10^{1} $& $5.7\times 10^{3}$\\

H-I &$1.1\times 10^{-10} $ & $9.7\times 10^{-7}$ & $ 9.1
\times 10^{-3}$& $8.8\times 10^{1} $& $1.1\times 10^{4}$\\
H-K &$2.2\times 10^{-11} $ & $2.1\times 10^{-7}$ & $ 2.1
\times 10^{-3}$& $2.1\times 10^{1} $& $2.7\times 10^{3}$\\
H-L &$6.7\times 10^{-11} $ & $6.1\times 10^{-6}$ & $ 5.7
\times 10^{-2}$&$5.4\times 10^{2} $ & $6.4\times 10^{4}$\\
H-V &$5.5\times 10^{-11} $ & $5.5\times 10^{-7}$ & $ 5.5
\times 10^{-3}$& $5.7\times 10^{1} $& $7.3\times 10^{3}$\\

I-K &$5.4\times 10^{-11} $ & $5.1\times 10^{-7}$ & $ 5.0
\times 10^{-3}$& $5.1\times 10^{1} $& $6.5\times 10^{3}$\\
I-L &$1.9\times 10^{-10} $ & $1.7\times 10^{-6}$ & $ 1.6
\times 10^{-2}$& $1.5\times 10^{2} $& $1.9\times 10^{4}$\\
I-V &$3.6\times 10^{-11} $ & $3.4\times 10^{-7}$ & $ 3.4
\times 10^{-3}$& $3.4\times 10^{1} $& $4.3\times 10^{3}$\\

K-L &$6.9\times 10^{-12} $ & $6.6\times 10^{-8}$ & $ 6.5
\times 10^{-4}$& $6.5\times 10^{0} $& $8.2\times 10^{2}$\\
K-V &$3.8\times 10^{-11} $ & $4.0\times 10^{-7}$ & $ 4.2
\times 10^{-3}$& $4.5\times 10^{1} $& $5.9\times 10^{3}$\\
L-V &$5.8\times 10^{-11} $ & $5.8\times 10^{-7}$ & $ 5.9
\times 10^{-3}$& $6.2\times 10^{1} $& $8.0\times 10^{3}$\\
\hline
\end{tabular}
\end{center}
\end{table}

\subsection{multiple pairs}
The squared optimal SNR of    multiple pairs  is  \cite{Allen}:
\ba\label{6 multiple pairs}
\rm SNR^2_{\rm  optimal} &=&
\rm {SNR^2_{AH}+ SNR^2_{AI}+ SNR^2_{AK}+ SNR^2_{AL}+ SNR^2_{AV} } \nonumber \\
&&  + \rm{ SNR^2_{HI}+SNR^2_{HK}+SNR^2_{HL}+SNR^2_{HV}+SNR^2_{IK} }  \\
&&  + \rm { SNR^2_{IL}+SNR^2_{IV}+SNR^2_{KL}+SNR^2_{KV}+SNR^2_{LV} } .   \nonumber
\ea
The result is shown in Table \ref{6SNR1},
which are three orders greater than
those of the multiple pairs of four detectors.
The corresponding minimum $\alpha_t$
is shown in Table \ref{6SNR1 constraint},
in particular,
RGW can be directly detected by the six 2nd-generation detectors
for   models with $\alpha_t>0.01364$ for   $(n_t=0,r=0.2)$.
Similarly, the minimum $n_t$ is calculated as
$ n_t>0.2982$ at $r=0.2$ and  $\alpha_t=0$.
This is a much improved capability
compared with   that of LIGO H-L S6
and of the four 1st-generation detectors.

\begin{table}[h]
\caption{\label{6SNR1}
SNR$_{\rm optimal}$  of the multiple pairs from six detectors at $ r=0.2$.
}
\begin{center}
\begin{tabular}{|c|c|c|c|c|c|}
\hline
$\alpha_t,~n_t$ & -0.005,~-0.1 & 0,~0 & 0.005,~0.1 & 0.01,~0.2 & 0.015,~0.2\\
\hline
\hline
optimal & $8.1\times 10^{-10} $ & $7.3\times 10^{-6}$&$ 6.8
\times 10^{-2} $ & $6.4\times 10^{2} $ & $7.8\times 10^{4}$\\
\hline
\end{tabular}
\end{center}
\end{table}

\begin{table}[h]
\caption{\label{6SNR1 constraint}
The minimum  $\alpha_t$ required to be detected
 by the multiple pairs from six detectors.}
\begin{center}
\begin{tabular}{|l|c|c|c|c|c|}
\hline
 $n_t=-0.2$ & $n_t=-0.1$ & $n_t=0$ & $n_t=0.1$ \\
\hline
\hline
$r=0.1$  & 0.01893 & 0.01437 & 0.00979 \\
$r=0.15$  & 0.01851 & 0.01394 & 0.00937 \\
$r=0.2$   & 0.01821 & 0.01364 & 0.00907 \\
$r=0.25$  & 0.01798 & 0.01341 & 0.00884 \\
$r=0.3$  & 0.01779 & 0.01322 & 0.00865 \\
\hline
\end{tabular}
\end{center}
\end{table}

\newpage
\subsection{six-detector correlation}
Similarly, the squared optimal SNR for six-detector correlation is  \cite{Allen}:
\ba
\rm  SNR^2_{\rm optimal} &=& \rm {SNR^2_{AH}SNR^2_{IK}SNR^2_{LV}+SNR^2_{AH}SNR^2_{IL}SNR^2_{KV}} \nonumber \\
&&+ \rm {SNR^2_{AH}SNR^2_{IV}SNR^2_{KL}+ SNR^2_{AI}SNR^2_{HK}SNR^2_{LV} }\nonumber \\
&&+ \rm {SNR^2_{AI}SNR^2_{HV}SNR^2_{KL}+SNR^2_{AI}SNR^2_{HV}SNR^2_{KL} } \nonumber \\
&&+ \rm {SNR^2_{AK}SNR^2_{HI}SNR^2_{LV}+SNR^2_{AK}SNR^2_{HL}SNR^2_{IV}} \nonumber \\
&&+ \rm { SNR^2_{AK}SNR^2_{HV}SNR^2_{IL}+SNR^2_{AL}SNR^2_{HI}SNR^2_{KV}} \\
&&+ \rm{SNR^2_{AL}SNR^2_{HK}SNR^2_{IV}+SNR^2_{AL}SNR^2_{HV}SNR^2_{IK} } \nonumber \\
&&+ \rm { SNR^2_{AV}SNR^2_{HI}SNR^2_{LK}+SNR^2_{AV}SNR^2_{HK}SNR^2_{IL} }\nonumber \\
&&+ \rm {SNR^2_{AV}SNR^2_{HV}SNR^2_{IK}},\nonumber
\ea
The calculated result is shown in Table \ref{6SNR2}.
One notices that,
in this case,
the optimal SNR$_{\rm optimal}$ is cubic in SNR$_{ij}$,
and its dependence on $(n_t,\alpha_t)$ is different from that of the multiple method.
For instance,
for the default  $(n_t=0,\alpha_t=0)$,
SNR$_{\rm optimal}$ of six-detector correlation
is smaller than SNR$_{\rm optimal}$ of multiple-pairs.
However, for greater values of $(n_t,\alpha_t)$,
SNR$_{\rm optimal}$ of six-detector correlation
trends to be much greater than SNR$_{\rm optimal}$ of multiple-pairs.
The minimum $\alpha_t$ for detection  are in Table \ref{6SNR2 constraint}.

\begin{table}[h]
\caption{\label{6SNR2}
SNR$_{\rm optimal}$ of  the six-detector correlation, at $r=0.2$.
}
\begin{center}
\begin{tabular}{|c|c|c|c|c|c|}
\hline
$\alpha_t,~n_t$ & -0.005,~-0.1 & 0,~0 & 0.005,~0.1 & 0.01,~0.2 & 0.015,~0.2\\
\hline
\hline
6 detectors & $5.6\times 10^{-30} $ & $4.5\times 10^{-18}$&$ 7.0
\times 10^{-6} $ & $3.7\times 10^{6} $ & $3.2\times 10^{12}$\\
\hline
\end{tabular}
\end{center}
\end{table}

\begin{table}[h]
\caption{\label{6SNR2 constraint}
The minimum  $\alpha_t$  by 6-detector correlation.}
\begin{center}
\begin{tabular}{|l|c|c|c|c|c|}
\hline
$n_t=-0.2$ & $n_t=-0.1$ & $n_t=0$ & $n_t=0.1$ \\
\hline
\hline
$r=0.1$  & 0.01958 & 0.014302 & 0.01046 \\
$r=0.15$  & 0.01916 & 0.01460 & 0.01004 \\
$r=0.2$  & 0.01886 & 0.01430 & 0.00974 \\
$r=0.25$  & 0.01863 & 0.01407 & 0.00951 \\
$r=0.3$ & 0.01844 & 0.01388 & 0.00932 \\
\hline
\end{tabular}
\end{center}
\end{table}

\subsection{How many detectors do we need?}

Here comes a question:  how many interferometers do we need to detect the RGW?
To a large extent, this depends on the parameters of RGW.
As an example, we can give an estimate for the multiple pairs method.
According to Eq.(5.46) in Ref.\cite{Allen},
for N detectors of the same level sensitivity,
one has
\be
{\rm SNR}^2_{\rm optimal}=\sum_{i<j }^{N} {\rm SNR}^2_{ij} \simeq  \frac{N(N-1)}{2} {\rm SNR}^2_{ij}.
\ee
By  $ {\rm SNR} _{ij}  \sim 10^{-6}$
for the default parameters  based on the second-generation detectors,
\be
 {\rm SNR_{optimal}} \simeq  10^{-6}\sqrt{\frac{N(N-1)}{2}}.
\ee
Then $N\simeq 10^6$ detectors would be required for ${\rm SNR} _{optimal}$
to be greater than $1$.
Nevertheless,
the chance to detect   RGW is largely dependent on
$n_t$ and $\alpha_t$.
For slightly greater values,
say $\alpha_t = 0.02$ and $n_t=0$ within the constraint of Eq.(\ref{constr2}),
six detectors will be enough to detect RGW
at a high significance with ${\rm SNR} \simeq 10^3$.

\section{ Conclusion}

Based on the analytical solution of RGW,
covering the whole five stages of cosmic expansion,
the resulting spectrum of RGW contains the three parameters  ($n_t,\alpha_t, r$),
that are determined by the initial condition during the inflation.
We take the amplitude of the spectrum
so that its low frequency end is equal to that of the primordial spectrum,
leading to a modification by $\sim 1/50$ to the previous result.

We have constrained ($n_t,\alpha_t, r$)
of the analytical spectrum,
using  the interferometer detectors.
The SNR of RGW are calculated
for various values of ($n_t$, $\alpha_t$, $r$),
based on the sensitivity data of S6 of LIGO HL pair.
The amplitude of  RGW
is very sensitive to    $\alpha_t$
at high frequencies $10^2-10^3$Hz.
We have derived the upper limits
$\alpha_t\le 0.02093$ at   $r=0.2$ and $n_t=0$,
and   $n_t<0.470$ at $(\alpha _t=0,r=0.2)$.
This is the first constraint upon $\alpha_t$ based on actual observations,
also consistent with the constraint on the energy density
obtained by LIGO-Virgo Collaboration.
CMB observations so far have little constraint on the indices $n_t$ and $\alpha_t$.
Thus, in regard to detections of RGW,
the interferometers method and the CMB observations are
complementary, working at different frequencies.

We have also explored the detection of RGW
by four 1st generation, and six 2nd generation detectors,
in two different optimal combinations.
We have calculated  SNR   in each case,
and  found that the chance to detect   RGW is largely dependent on
$n_t$ and $\alpha_t$ of RGW itself.
RGW can be directly detected
by the six 2nd-generation detectors
for   models with $\alpha_t>0.01364$ at $r=0.2$ and $n_t=0$,
or $ n_t>0.2982$ at $r=0.2$ and  $\alpha_t=0$.

\

We thank Dr. Nishizawa for helpful information and discussions.
Y. Zhang is supported by NSFC Grant No. 11275187, NSFC 11421303,
SRFDP, and CAS,
the Strategic Priority Research Program
"The Emergence of Cosmological Structures"
of the Chinese Academy of Sciences, Grant No. XDB09000000.

\

\

\appendix

{\Huge \bf Appendix}

\

\begin{figure}[h]
\begin{center}
\includegraphics[width=0.6\textwidth]{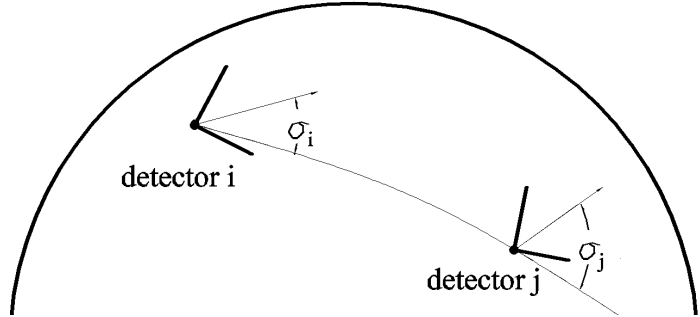}
\caption{\label{ij}
The angles $\sigma_i$ and $\sigma_j$ are schematically shown
for two detectors $i$ and $j$ on the Earth.}
\end{center}
\end{figure}

In the expression of SNR  in Eq.(\ref{SNR}),
the overlapping function $\gamma(f)$ between a pair of two detectors $i$ and $j$
appears.
To calculate $\gamma(f)$,
one needs to know  the positions and directions of the two detectors in the pair.
The overlapping function between the detectors $i$ and $j$
 is defined  by \cite{Flanagan, Allen,  Japanese}:
\ba
\gamma(f)=  \Theta _{+}(\alpha,\, \beta)\,\cos(4\sigma_{+})
            + \Theta _{-}(\alpha,\, \beta)\,\cos(4\sigma_{-}) \;,
\label{gamma}
\ea
where
\ba
\Theta _{+} (\alpha,\,\beta) &\equiv & - \left( \frac{3}{8} j_0 -\frac{45}{56} j_2 + \frac{169}{896} j_4 \right)
+ \left( \frac{1}{2} j_0 -\frac{5}{7} j_2 - \frac{27}{224} j_4 \right) \cos \beta \nonumber \\
&&- \left( \frac{1}{8} j_0 +\frac{5}{56} j_2 + \frac{3}{896} j_4 \right) \cos 2\beta \;, \nonumber \\
&& \nonumber \\
\Theta _{-} (\alpha,\,\beta) &\equiv & \left( j_0 +\frac{5}{7} j_2 + \frac{3}{112} j_4 \right) \cos \left( \frac{\beta}{2} \right)^4 \;\nonumber,
\label{eq33}
\ea
consisting of $j_n=j_n(\alpha)$ is the spherical Bessel's functions.
In the above,
 $\beta $ is the angle formed by the two detectors
measured at the center of the Earth,
$\alpha$ is the phase difference between the two detectors
due to the separation $|\Delta \vec{X}|$ of  the two detectors:
\[
\alpha(f)=2\pi f  \frac{|\Delta \vec{X}|}{c} , \,\,\,\,\,\,\,\,\,
|\Delta \vec{X}| = |\vec{X}_i -  \vec{X}_j|=2R_E\sin \frac{\beta}{2},
\]
with $R_E=6371$km being the radius of Earth.
In Eq.(\ref{gamma}),
\[
\sigma_+=  \frac{1}{2}(\sigma_i+\sigma_j), \,\,\,\,\,\,\,\,\,
\sigma_-=  \frac{1}{2}(\sigma_i-\sigma_j),
\]
where $\sigma_i $
is the angle between the tangent line of the great circle connecting the two detectors
and the bisector between two arms of the detector $i$,
and $\sigma_j$ is for $j$, schematically shown in Fig \ref{ij}.

In this paper,
 seven detectors are considered,
AIGO(A), Geo(G), LIGO-Hanford(H), LIGO-India(I), Kagra(K), LIGO-Livingston(L), and Virgo(V).
Information of the positions and directions of
seven detectors is listed in   Table \ref{detectors}
\cite{Japanese}, \cite{Seto}, \cite{LIGO-India}.
The  position of a detector  on  the  Earth
is fixed by $\theta$ and $\phi$ in the spherical coordinate.
$\psi$  is the angle  between the local east direction
and the bisecting line of two arms measured counterclockwise.
From these given values,
we have calculated  $(\beta, \sigma_+, \sigma_-)$ of the relevant pairs,
and the result is given in Tab.\ref{positions and orientations}.
In Ref. \cite{Japanese}, some of the values of $\sigma_+$ and $\sigma_- $
are different from ours in Table \ref{positions and orientations}.
Ref.  \cite{Japanese} has adjusted the values of  $\sigma_+$ and $\sigma_-$
by an integer times of $90^{\circ}$.
This brings no difference to $\cos(4\sigma_{+})$ and $ \cos(4\sigma_{-})$
that appear in the overlapping function $\gamma(f)$.

\begin{table}[h]
\caption{\label{detectors}
Positions and orientations of seven interferometers on the Earth  (in units of degree)}
\begin{center}
\begin{tabular}{|c|c|c|c|c|}
\hline
detector & $\theta$ & $\phi$  & $\psi$\\
\hline
\hline
 $A$ & $121.4$ & $115.7$ & $-45.0$ \\
 $G$ & $47.7$ & $9.8$ & $68.8$ \\
 $H$ & $43.5$ & $-119.4$ & $171.8$ \\
 $I$ & $70.9$ & $74.05$ & $135.0$ \\
 $K$ & $35.6$ & $137.3$ & $70.0$ \\
 $L$ & $59.4$ & $-90.8$ & $243.0$ \\
 $V$ & $46.4$ & $10.5$ & $116.5$ \\
\hline
\end{tabular}
\end{center}
\end{table}

\begin{table}[h]
\caption{\label{positions and orientations}
Angles  and separations appearing in $\gamma(f)$ for the relevant pairs}
\begin{center}
\begin{tabular}{|c|c|c|c|c|}
\hline
detector pair & $\beta$ ($^{\circ}$)& $|\Delta \vec{X}| (km) $
        & $\sigma_+$ ($^{\circ}$)& $\sigma_-$($^{\circ}$)\\
\hline
\hline
 $A-H$ & $135.6$ & $1.18\times 10^{4}$ & $-44.9$ & $-36.3$ \\
 $A-I$ & $64.4$ & $6.79\times 10^{3}$ & $3.4$ & $-182.6$ \\
 $A-K$ & $70.8$ & $7.38\times 10^{3}$  & $-148.6$ & $31.9$ \\
 $A-L$ & $157.3$ & $1.25\times 10^{4}$ & $2.1$ & $-52.0$ \\
 $A-V$ & $121.4$ & $1.11\times 10^{4}$ & $-29.2$ & $-160.8$ \\
 $G-H$ & $80.4$ & $8.23\times 10^{3}$ & $31.7$ & $-85.6$ \\
 $G-L$ & $77.0$ & $7.93\times 10^{3}$ & $60.0$ & $-141.4$ \\
 $G-V$ & $1.4$ & $1.55\times 10^{2}$ & $-65.8$ & $65.9$ \\
 $H-I$ & $113.3$ & $1.06\times 10^{4}$ & $61.5$ & $6.5$ \\
 $H-K$ & $72.4$ & $7.52\times 10^{3}$ & $25.6$ & $0.9$ \\
 $H-L$ & $27.2$ & $3.00\times 10^{3}$ & $152.2$ & $45.3$ \\
 $H-V$ & $79.6$ & $8.16\times 10^{3}$ & $55.1$ & $66.1$ \\
 $I-K$ & $57.5$ & $6.13\times 10^{3}$ & $-2.9$ & $106.3$ \\
 $I-L$ & $128.2$ & $1.15\times 10^{4}$ & $99.8$ & $-71.5$ \\
 $I-V$ & $58.0$ & $6.18\times 10^{3}$ & $57.6$ & $-62.5$ \\
 $K-L$ & $99.2$ & $9.71\times 10^{3}$ & $68.1$ & $-47.6$ \\
 $K-V$ & $86.6$ & $8.74\times 10^{3}$ & $5.6$ & $-61.1$ \\
 $L-V$ & $76.8$ & $7.91\times 10^{3}$ & $83.1$ & $116.7$ \\
\hline
\end{tabular}
\end{center}
\end{table}

\end{document}